# Formation and characterization of an Al-rich metastable phase in Al-B phase diagram

Alexander I. Malkin[1], Vladimir V. Chernyshev[1,2], Alena A. Ryazantseva[1],

Alexander L. Vasiliev[3,4], Maximilian S. Nickolsky[5], Andrei A. Shiryaev[1],*

[1] Frumkin Institute of physical chemistry and electrochemistry, Russian Academy of Sciences, Leninsky Pr. 31. korp 4, 119071 Moscow, Russia; a_shiryaev@mail.ru

[2] Department of Chemistry, Lomonosov Moscow State University, Leninskie Gory 1, bld. 3, 119991 Moscow, Russia; vvc1955@yandex.ru

[3] Shubnikov Institute of Crystallography of Federal Scientific Research Centre "Crystallography and Photonics" of Russian Academy of Sciences, Leninskii Prospekt 59, Moscow, 119333, Russian Federation; a.vasiliev56@gmail.com

[4] Moscow Institute of Physics and Technology, 9 Institutskiy per., Dolgoprudny, Moscow region, 141701, Russian Federation

[5] Institute of Ore Geology, Petrography, Mineralogy and Geochemistry (IGEM), Russian Academy of Sciences, Staromonetny per, 35, 119017 Moscow, Russia; mnickolsky@gmail.com

\* - Corresponding author: Andrei A. Shiryaev, a_shiryaev@mail.ru

**Abstract**

Vacuum heat treatment of mechanically-alloyed powders of boron and aluminum leads to formation of a metastable Al-rich phase, which can be quenched. Its structure, composition, and thermal stability are established. With chemical formula $Al_{1.28}B$ the rhombohedral phase is unusually rich in Al. Parameters of the unit cell determined from X-ray powder diffraction are: a=18.3464(19) Å, c=8.9241(9) Å, 2601.3(6) Å$^3$, space group $R$-3. It is stable at heating up to 630 °C. It is suggested that this phase is an important intermediate step in formation of $AlB_2$ and, eventually, of other borides and its nucleation and thermal stability are explained by high elastic energy hindering formation of equilibrium phases at low temperatures.

1. Introduction

Boron-aluminum compounds and composite materials are of great interest for various applications. It is well known, that alloying of aluminum with boron and boron compounds promotes grain refinement, precipitation hardening and mechanical reinforcement of aluminum-based alloys. Boron-aluminum composite materials are used, in particular, for nuclear and space applications (Harrigan, 2018). The use of boron-aluminum compounds is also considered as a promising way to create new energetic materials (Whittaker et al., 2012).

Despite practical significance and numerous studies, controversies in equilibrium boron-aluminum phase diagram still exist, especially in B-rich field (Duschanek and Rogl, 1994; Mirkovic et al., 2004). All known versions, nevertheless, share a common trait, namely, the absence of stable intermetallic phases with Al:B ratio less than 1:2, i.e. $AlB_2$. As for the metastable boron-aluminum compounds, their mere existence and structure remain debatable. The only work reporting an Al-rich boron-aluminum phase is that by R. Vardiman (1992). The novel rhombohedral phase with a boron content between 57-62 at% has been formed upon annealing a strongly supersaturated solid solution obtained by boron ion implantation in aluminum. The phase is most likely metastable, but yet is tolerant to vacuum heat treatment to at least 525 °C. It was suggested that the detected phase is nucleated at structure defects of the aluminum matrix stabilized by boron atoms. That assumption seems natural but does not explain the amazing stability of the new phase.

Metastable intermetallic phases are usually produced by various methods of mechanical alloying (Han et al., 2019; Suryanarayana et al., 2011; Zhilyaev and Langdon, 2008; Zhang et al., 1994) or high rate cooling (Kube and Shroers, 2020; Wei et al., 2018). The structure and stability limits of metastable phases are therewith governed by the fabrication procedure. In regard to boron-aluminum system, it is important that the solubility of boron in aluminum is extremely low (Duschanek and Rogl, 1994) Therefore, formation of intermetallic compound by decomposition of supersaturated solid solution seems to be a very exotic process. Facts discussed above raise several important questions concerning the Al-B phase discovered in (Vardiman, 1992). First, why this phase was not observed in numerous works on the synthesis of aluminum borides by traditional methods? Second, is it possible that this phase appears in other processes not associated with the annealing of highly supersaturated solutions? Finally, what are the reasons for its persistence at elevated temperatures and what are the conditions of its formation and decomposition? In addition, the composition of the new phase was determined rather approximately and therefore needs to be refined.

Here we consider phase transformations in boron-aluminum composite powders obtained by mechanical alloying. The formation of rhombohedral aluminum-rich phase (for sake of brevity termed an "M-phase", "M" stands for "metastable") at vacuum heat treatment is observed. Atomic structure, chemical composition the conditions of formation and decomposition of this phases are established. Driving force and mechanisms of its formation and of thermodynamic stability are discussed.

2. **Material and methods**

The composite powder was produced from the mixture of aluminum and fine boron with boron content 45 mass % (66 at. %) by mechanical alloying in a water-cooled ball mill in hexane medium. The resulting product contained composite particles with rather uniform distribution of submicron boron particles in the aluminum matrix and relatively small amount of free boron not embedded in aluminum. Volatile impurities in the composite powder were removed by vacuum sublimation at 120 °C. Content of residual impurities (Si, C, N, O) was about 1.2 mass %. The boron-aluminum specific interface area was about 1.8 $m^2/g$ with an average volumetric size of the composite particles of 43±2 μm. The detailed data on the structure and constitution of composite powders are presented in (Malkin et al., 2019).

The phase composition was studied *in situ* using high temperature X-ray diffraction using Empyrean (Panalytical BV) diffractometer equipped with an HTK-1200N oven (Anton Paar) and 1D position sensitive detector X'Celerator. The system operates in Bragg-Brentano (reflection) geometry with Ni-filtered Cu Kα-radiation. The height of the sample in the oven is automatically adjusted to compensate thermal expansion. The experiments were performed in vacuum with different ramp rates and temperature steps. However, at least qualitatively, the results are only temperature-dependent, i.e., experimental scheme do not play a prominent role. The ramp rate was 20 °/min, the maximal temperature was 700 °C. Pattern acquisition time at a given temperature step was always less than 25 minutes. In the kinetic experiments, only the most representative angular ranges were studied and duration of individual scans was less than 2 minutes. Cooling was performed by switching the oven power. The structure of the M-phase was solved from high resolution scans performed on a heat-treated sample after quenching to room temperature and after removal of excess unreacted boron by ethanol. The structure was solved from powder diffraction data; since this approach is less common than single crystal data evaluation, details of the procedure are given in respective part of the paper and in Supplementary materials.

To evaluate chemical composition of the M-phase Scanning and Transmission Electron microscopies (SEM and TEM) were employed. For TEM the powder of a sample heated to 500 °C and with removed excess boron was embedded in epoxy resin and cured at a temperature of 80 °C. The obtained specimen were ground mechanically to the thickness of 50-100 μm and thinned by $Ar^+$ ions with energy of 5 keV at the beginning and 1 keV at the final step in the Gatan PIPS (Gatan, USA) until small holes appeared. Microstructural analyses were performed in a OSIRIS TEM/STEM (Thermo Fisher Scientific, USA) equipped with a high angle annular (HAADF) electron detector (Fischione, USA) and Bruker energy-dispersive X-ray microanalysis system (Bruker, USA) at an accelerating voltage of 200 kV. Image processing was performed using a Digital Micrograph (Gatan, USA) and TIA (ThermoFisher Scientific, USA) software. Selection of the grains for the analysis was made based on their diffraction patterns.

3. **Results**

   3.1. High temperature diffraction

   X-ray diffraction pattern of the sample measured in vacuum at 570 °C is shown in Fig. 1; patterns recorded in the temperature range 25-700 °C are shown in Fig. S1 (see Supplementary Materials). The initial sample shows strong peaks from metal aluminum and a broad halo from amorphous boron. Upon heating, at 470-480 °C, in addition to aluminum, a new phase (the M-phase) is observed. The formation of the M-phase appears to be rather fast, since upon reaching the preset temperature of 470 °C it was already present. The exact temperature of M-phase nucleation remains uncertain, but it likely depends on the ramp rate. Besides temperature-induced lattice expansion, no structural changes of the M-phase were observed in the whole range of its stability. At 630 °C the peaks of this phase rapidly decrease in intensity and $AlB_2$ phase appears. At 640 °C the M-phase fully disappears on time scale of ~5 min with corresponding increase of the $AlB_2$ content. The subsequent heating up to 700 °C does not lead to other changes. Note that, upon cooling from 700 °C to room temperature, the M-phase does not reappear and amount of $AlB_2$ does not change. Experiments in air do not lead to formation of the M-phase, instead an orthorhombic aluminum borate $Al_4B_2O_9$ emerges.

   3.2. Structure and composition of the Al-rich phase

   Since unambiguous deciphering of the structure of the new phase requires knowledge of its chemical composition, results of the X-ray diffraction and electron microscopy are discussed simultaneously. Whereas it was easy to remove excess unreacted boron by washing in ethanol, an attempt to remove Al metal by acids from the quenched sample failed, since both Al and the new phase were dissolved. X-ray powder diffraction pattern measured at 570 °C *in vacuo* was used for initial indexing and structure solution, and powder pattern measured at 25 °C after the quenching and boron removal was used for the M-phase crystal structure completion and final two-phase refinement. In both cases, sample contained two crystalline phases: Al metal and M-phase. For additional detail of the structure solution see Supplementary materials.

In the high-temperature pattern (T = 570 °C), the positions of twenty one diffraction peaks from the M-phase (see Table S1) were indexed using TREOR90 (Werner et al., 1985) and AUTOX (Zlokazov 1992, 1995) software in hexagonal R-centered lattice with the following parameters: $a$ = 18.46 Å and $c$ = 8.98 Å. Application of the Pawley fitting (Pawley, 1981) narrowed the search to the following five space groups (s.g.): *R*3, *R*-3, *R*32, *R*3*m* and *R*-3*m* and, in addition, showed that 111 and 200 reflections from the Al metal do not overlap with signal of the new phase in the high temperature pattern.

Further steps required knowledge of chemical composition of M-phase. Analysis using SEM equipped with an EDX detector with Be window ruled out presence of noticeable C, N, O content. However, due to inevitable contribution of several phases, SEM-EDX data required independent confirmation, which was achieved by TEM-EDX. A representative high resolution image of a grain of the M-phase with corresponding FT-image is shown in Fig. 2. Analysis of several independent grains gave atomic composition of the phase as 45 at.% B and 55 at.% Al with an estimated uncertainty of ±5 at%; these values are close to those obtained from SEM.

Using constrained chemical composition, initial search of the structural motif of the new phase was performed using simulated annealing method realized in MRIA (Zlokazov, Chernyshev, 1992) and FOX (Popa, 1998) software. Rietveld refinement of all motifs found using MRIA was performed; isotropic thermal vibrations were also refined at an advanced step. Subsequent crystal chemistry analysis narrowed the solutions to those with Al–Al, Al–B and B–B interatomic distances not less than 2.4, 2.0 and 1.8 Å, respectively. The best variants belong to two space groups *R*-3 and *R*-3*m* only. They were used as the starting ones in Rietveld refinement of the room temperature pattern, and this refinement gave preference to the only space group *R*-3. Consequently, analysis of difference Fourier maps was performed to locate possible additional boron sites. A boron atom was placed into every electron density maximum stronger than 1 e/Å$^3$ and its position and occupancy were refined. This procedure allowed to localize an additional boron site on the 3rd order axis. On the following steps of the Rietveld refinement coordinates of all atoms and a common isotropic thermal vibration parameter $U_{iso}$ were analysed. This allowed to determine the site occupancies and to reveal the sites, which can be occupied by both Al and B.

In the final two-phase Rietveld the following parameters of the M-phase were refined: the cell parameters *a* and *c*; peak profiles accounting for anisotropic line broadening (Popa, 1998); coordinates of 5 Al and 4 B sites with occupancy of 1; coordinates of two sites occupied by Al and B with refinement of the occupancies; coordinate and occupancy of the (0, 0, z) boron site; common isotropic thermal parameter $U_{iso}$. The refinement was stopped at $\chi^2 = 2.362$ (Fig. 3). The main crystallographic data are shown in Table 1; atomic coordinates are given in Table 2 and CIF file (Cambridge Structure Database entry No. CSD 2069017). Taking into account the data from the tables, the ratio of Al:B in the new phase is equal to 1.28. The obtained structure is shown in Fig. 4.

We note that although the Al:B ratio of 1.28 is the preferred one, the occupancy of the boron (0, 0, z) site may vary slightly. However, in any case the Al:B ratio should be >1.2. Interestingly, almost all diffraction peaks of the new phase could be fitted also in the s.g. *R-3c*; in such a case the B12 position (i.e. a partially occupied position of boron in a channel) is empty. However, in this s.g. the peaks at 29.76 and 29.94° should be absent. In reality, these peaks are always observed in all acquired patterns and, consequently, the eventual end member without the B12 atom does not exist.

Analysis of the patterns acquired at 25, 500 and 570 °C allows estimation of the thermal expansion coefficient of the new phase, which is found to be $\sim 3.3 \times 10^{-5}$ K$^{-1}$; the value similar to that calculated for AlB$_2$ (Xiao-Lin et al., 2006).

**4. Discussion**

The structure of the M-phase is identical to the one reported by R. Vardiman (1992); however, Al:B ratio differ significantly from previously reported. According to (Vardiman, 1992), the boron content falls in range 57-62 at %, corresponding to formulas such as $Al_3B_4$, $Al_2B_3$ or $Al_5B_8$. However, an attempt to fit these compositions to the solved structure of the M-phase failed. Presumably, the Auger analysis in (Vardiman, 1992) was compromised by incorrect evaluation of contribution of surrounding aluminum matrix due to small size of the precipitates. The data reported by Vardiman (1992) are nevertheless very useful. Only Al and B were detected in measurable quantities both in (Vardiman, 1992) and in our work, ruling out eventual presence of O, N, C and Si. Moreover, the formation of the M-phase in markedly different conditions gives an insight into its nucleation and growth.

The formation of metastable M-phase is obviously governed by kinetic restrictions on the formation of equilibrium borides. Otherwise, boron particles should have been enveloped into a multi-shell structure with a sequence of phases $AlB_{12}$, $AlB_2$ and M. The question is what are the reasons for the kinetic limitation? The kinetics of intermetallides formation generally depends on the diffusivities in individual phases and the rate of atoms transfer across interface. Let us first assume that boron diffusion is the rate-determining process. While the multi-shelled structure has been formed, the growth rate of each phase depends on permeability of adjacent ones (Geguzin, 1979). In this connection it should be noted that, unlike $AlB_2$ and $AlB_{12}$, the M-phase lattice possess wide channels along the c-axis (see fig. 4). It is thus plausible that the boron diffusivity in the M-phase should therefore be much greater than in equilibrium phases. Presumably, at least at intermediate stages of phase transformation, the flux of boron atoms to «$AlB_2$ – M-phase» interface cannot compensate their outflow to the outer boundary of M-phase. In other words, the layers of equilibrium phases, if formed previously, should thin and eventually disappear, thus implying kinetic prohibition of their formation. However, the increase in M-phase thickness clearly implies decrease in the boron outflow from its inner boundary. Because of this, a limiting layer thickness at which the prohibition disappears should exist. The above considered experimental data indicate that the limiting thickness in quite sizeable. Note also that the diffusion flux through the M-phase has to be strongly anisotropic, so that the phase growth should predominantly occur along c-axis.

At the initial stage of phase formation, while the total thickness of the resulting product layer is still small, the rate-determining process is boron transfer across the interface. Even if equilibrium phases come into existence at this stage, they should disappear in the future. However, it is most likely that the M-phase is formed precisely at the boron-aluminum interface. An argument for this assumption is the result of work (Vardiman, 1992) where direct formation of the M-phase from supersaturated solid solution of boron in aluminum was observed.

Decomposition of supersaturated solid solution can be represented as a competitive nucleation and growth of supercritical nuclei of new phases. As is well known, the rate of formation of supercritical nucleus depends not only on the chemical potentials of the components in the initial phases but also on the strain energy arising from the difference in specific volumes of phases (the volume effect), and the specific interfacial energy (Lifshitz, Slyozov, 1961). We note that formation of the M-phase was observed already after annealing at 200 °C only (Vardiman, 1992), which is a very low value. Since the heat of formation of the M-phase is obviously less than that of $AlB_2$ and $AlB_{12}$, the absence of equilibrium phases seems to be the result of contribution of the strain energy and, to a lesser extent, the interfacial energy to the work of formation of the nuclei of critical size. Note that volume effect in the formation of $AlB_2$ and $AlB_{12}$ is much greater than that of M-phase. In the synthesis of aluminum borides from the elements, the relative volume effects are −0.198, −0.063 and −0.028 for unstrained $AlB_2$, $AlB_{12}$ and M-phase, respectively. Consequently, the strain energy, which is proportional to volume effect squared, at the M-phase nucleation at the Al-B interface is much less than in case of equilibrium borides.

The stability of the M-phase at temperatures up to 630 °C is explained by high energy barrier of phase transformation «M-phase → Al + $AlB_2$». At temperatures close to the melting point of aluminum, the stress relaxation rate becomes so high, that the kinetic restrictions on the formation of equilibrium phases disappear. These considerations are also applicable to the synthesis of aluminum borides by traditional high-temperature methods.

## 5. Conclusions

Phase transformations in powders of mechanically alloyed boron-aluminum composite are considered. Vacuum heat treatment produces easily quenchable metastable phase, which is stable up to 630 °C. The lowest temperature of its formation lies between 200 and 470 °C, likely being a function of annealing and synthesis conditions. This phase is identical to the one obtained by annealing of B-implanted Al metal. Structure of this phase is solved giving a unit cell with a = 18.3464(19) Å, c = 8.9241(9) Å, V = 2601.2(6) Å$^3$ and space group *R*-3. The phase is unusually Al-rich with formula Al$_{1.28}$B.

This work shows that formation of the Al-rich metastable phase with composition close to Al$_{1.28}$B is not restricted to rather exotic process such as ion implantation, but can be achieved by more common methods, including mechanical alloying. Therefore, investigation of its properties and behavior is of broad interest. Formation of this phase and its thermal stability are explained by very high elastic energy required for nucleation of equilibrium phases.

**Supplementary Materials.**

Additional details of the structure evaluation procedure, HT-XRD patterns, CIF file and atomic coordinates for the new phase are provided as Supplementary materials.

**Acknowledgements**.

X-ray diffraction experiments were performed using equipment of Center of Shared Use of IPCE RAS. We thank two anonymous reviewers for useful comments.

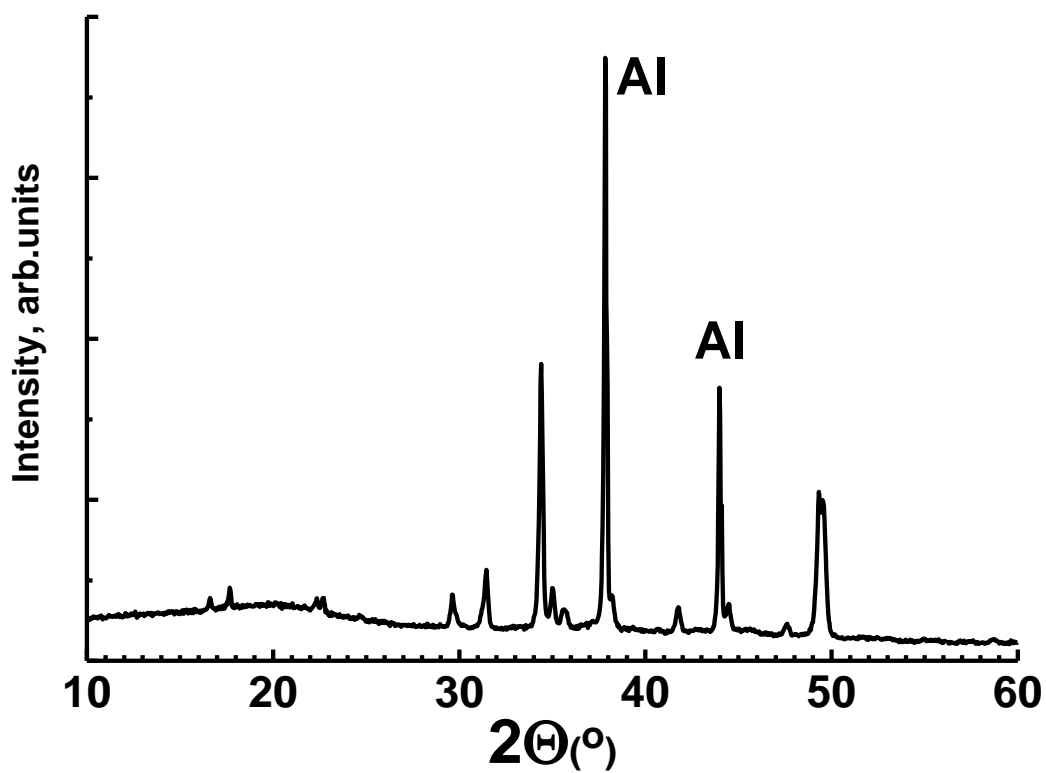

**Figure 1.** X-ray diffraction pattern of the Al-B mixture recorded at 570 °C in vacuum. Positions of reflexes of residual Al metal are shown.

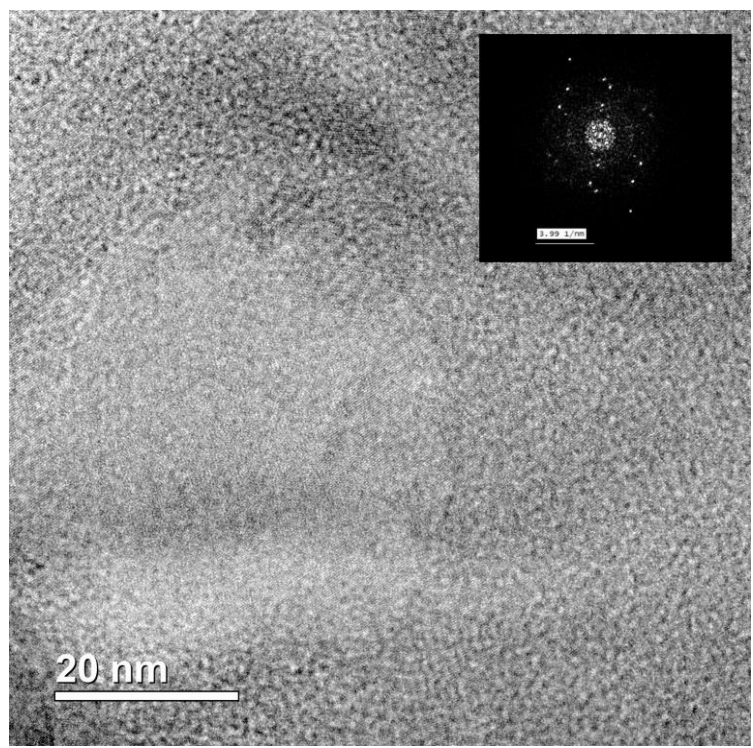

**Figure 2.** High-resolution TEM image of a grain of a new phase and its FFT pattern. Main peaks at the FT correspond to distances of 1.8, 2.6 and 3.0 Å (see also table S1).

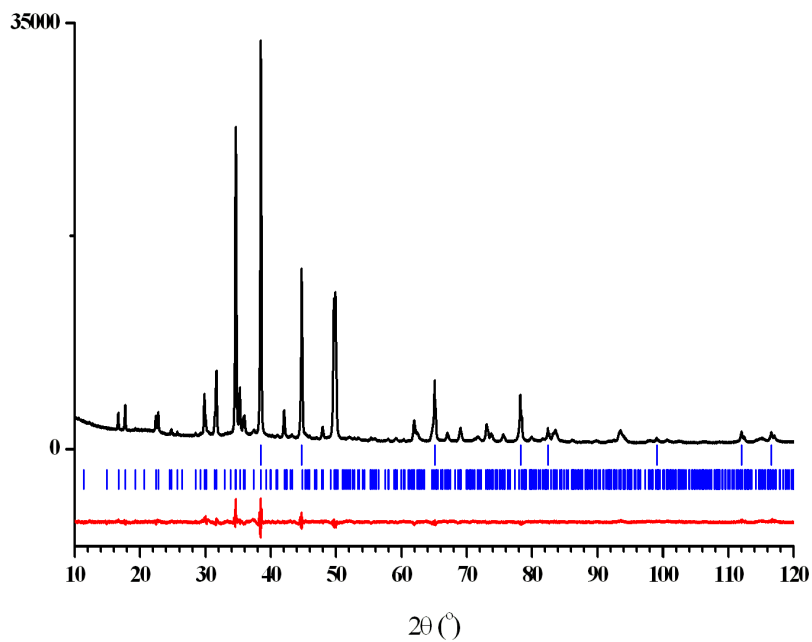

**Figure 3.** Rietveld plot of RT X-ray pattern showing the experimental (black) and difference (red) curves after two-phase refinement. Vertical bars denote calculated positions of the peaks of cubic **Al** (1st row) and **M** (2nd row) crystalline phases.

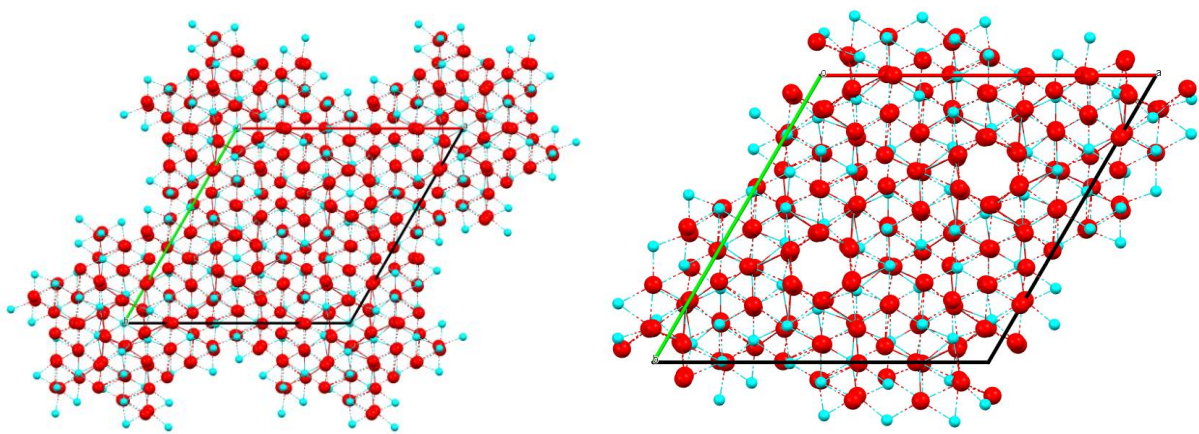

**Figure 4.** Crystal lattice of the M-phase ($Al_{1.28}B$). Left - Projection along the c axis. Right – projection with the B12 atom removed. Al atoms are red, B – cyan.

**Table 1. Crystallographic data for the M-phase.**

| | |
|---|---|
| CSD number | 2069017 |
| Formula | $Al_{1.28}B$ |
| Crystal system | Rhombohedral/Hexagonal |
| Space group | $R$-3 |
| a (Å) | 18.3464(19) |
| c (Å) | 8.9241(9) |
| V (Å$^3$) | 2601.2(6) |
| $M_{20}^*$ | 15 |
| $F_{21}^*$ | 16 (0.011, 35) |
| Z | 18 |
| Density, g/cm$^3$ | 2.570 |
| Diffractometer | EMPYREAN |
| Radiation, wavelength (Å) | CuK$_\alpha$, 1.5418 |
| Diffraction pattern: $2\theta_{min}$ - $2\theta_{max}$, step (°) | 10.000 – 120.000, 0.008 |
| $R_p/R_{wp}/R_{exp}/\chi^{2*}$ | 0.033/0.043/0.026/2.362 |

\* - Markers of the indexing quality $M_{20}$ and $F_{21}$ are defined in (de Wolff, 1968) and (Smith and Snyder, 1979), respectively. $R_p$, $R_{wp}$, $R_{exp}$ and $\chi^2$ defined in (Young and Wiles, 1982).

**Table 2. Atomic positions and occupancies in M-phase. Common isotropic displacement parameter $U_{iso}$ was refined to 0.033(2) Å$^2$.**

| Atom | x | y | z | Occupancy factor (s.o.f.) |
|------|---|---|---|---------------------------|
| Al1 | 0.2721 | 0.1289 | 0.0570 | 1 |
| Al2/B2 | 0.1186 | 0.0576 | 0.4165 | 0.69(2)/0.31(2) |
| Al3 | 0.3368 | 0.0557 | 0.3946 | 1 |
| Al4 | 0.1952 | -0.0044 | 0.2497 | 1 |
| Al5 | 0.7832 | -0.0002 | 0.2437 | 1 |
| Al6 | 0.1349 | 0.2697 | 0.4439 | 1 |
| Al7/B7 | 0.0705 | 0.1168 | 0.0902 | 0.62(2)/0.38(2) |
| B8 | 0.3293 | 0.0721 | 0.1634 | 1 |
| B9 | 0.2020 | 0.7981 | 0.4932 | 1 |
| B10 | 0.2043 | -0.0027 | 0.4817 | 1 |
| B11 | -0.0833 | 0.0641 | 0.1475 | 1 |
| B12* | 0 | 0 | 0.3547 | 0.76(3) |

\* - The B12 atom is on the 3-rd order axis. In the *R-3* space group multiplicity of this site is 1/3.